\documentclass[cameraready]{Interspeech}
\usepackage[utf8]{inputenc} 
\usepackage[T1]{fontenc}    
\usepackage{hyperref}       
\usepackage{url}            
\usepackage{amsfonts}       
\usepackage{nicefrac}       
\usepackage{microtype}      
\usepackage{booktabs}
\usepackage{adjustbox}
\usepackage{array}
\usepackage{subcaption}
\usepackage{fontawesome5}
\usepackage{tabularx,booktabs}
\usepackage{amsmath}
\usepackage{amssymb}
\usepackage{booktabs}
\usepackage{graphicx}

\usepackage{rotating} 

\usepackage{multirow}
\usepackage[table]{xcolor} 
\usepackage{cleveref}

\definecolor{xiaomiblue}{HTML}{4A7BCE}      
\definecolor{xiaomipaleblue}{HTML}{B8DCFE}  
\definecolor{xiaomiorange}{HTML}{FFA903}    
\definecolor{xiaomiteal}{HTML}{03CCA0}      
\definecolor{xiaomigreen}{HTML}{50B341}     
\definecolor{xiaomicoral}{HTML}{ED696D}     
\definecolor{xiaomilightgray}{HTML}{AAAAA8} 
\definecolor{xiaomibrightblue}{HTML}{04A3FD}
\definecolor{xiaomimedgray}{HTML}{6E6E6C}   
\definecolor{xiaomiblack}{HTML}{030303}     
\definecolor{xiaomired}{HTML}{ee4028}

\title{The Interspeech 2026 Audio Encoder Capability Challenge for Large Audio Language Models}

\author[affiliation={1}]{Heinrich}{Dinkel}
\author[affiliation={1}]{Jiahao}{Zhou}
\author[affiliation={2}]{Guanbo}{Wang}
\author[affiliation={1}]{Yadong}{Niu}
\author[affiliation={1}]{Junbo}{Zhang}
\author[affiliation={2}]{Yufeng}{Hao}
\author[affiliation={2}]{Ying}{Liu}
\author[affiliation={2}]{Ke}{Li}
\author[affiliation={3}]{Wenwu}{Wang}
\author[affiliation={4}]{Zhiyong}{Wu}
\author[affiliation={1}]{Jian}{Luan}

\address{
    $^1$ MiLM Plus, Xiaomi Inc., Beijing, China\\ 
    $^2$ DataoceanAI Inc., USA\\
    $^3$ Centre for Vision Speech and Signal Processing, University of Surrey, Guildford, UK\\
    $^4$ Shenzhen International Graduate School, Tsinghua University, Shenzhen, China
}

\email{\{dinkelheinrich, zhangjunbo1\}@xiaomi.com}


\usepackage{comment}

\begin{document}

\maketitle

\begin{abstract}
This paper presents the Interspeech 2026 Audio Encoder Capability Challenge, a benchmark specifically designed to evaluate and advance the performance of pre-trained audio encoders as front-end modules for Large Audio Language Models (LALMs). 
While LALMs have shown remarkable understanding of complex acoustic scenes, their performance depends on the semantic richness of the underlying audio encoder representations. 
This challenge addresses the integration gap by providing a unified generative evaluation framework, XARES-LLM, which assesses submitted encoders across a diverse suite of downstream classification and generation tasks.
By decoupling encoder development from LLM fine-tuning, the challenge establishes a standardized protocol for general-purpose audio representations that can effectively be used for the next generation of multimodal language models.
\end{abstract}

\section{Introduction}

The emergence of Large Audio Language Models (LALMs) has shifted the frontier of audio processing from simple single-task classification to complex unified understanding. 
Unlike previous classification models, modern LALMs offer a more intuitive user experience by generating natural language explanations rather than traditional categorical labels.
However, architectural diversity remains a bottleneck, since most state-of-the-art LALMs~\cite{chu2024qwen2,dinkel2025midashenglm,audioflamingo} rely exclusively on a limited selection of pretrained audio encoders, mainly based on Whisper. 
To address this issue we propose the Interspeech 2026 Audio Encoder Capability Challenge, aiming to assess a wide variety of possible audio encoders as LALM front-ends.

Different from the majority of previous benchmarks, we assess audio encoder performance using \textit{a single} decoder model in the form of a large language model (LLM).
In practice, this means that compared to previous challenges, which output logits and probabilities, our challenge requires human-readable outputs.
The challenge is structured into four different tracks: Track A, Track A (Hidden), Track B and Track B (Hidden).
Participants may use the open-source baseline framework to evaluate their performance on Track A and Track B, which can both be seen as development datasets.
Track A (Hidden) and B (Hidden) serve as evaluation tracks, which are unknown to participants.

Across these four tracks, in three of which we train a single dedicated LLM decoder, as seen in \Cref{tab:task_structure}.
Only for Track B (Hidden) we use the LLM decoder previously trained on Track B for evaluation.
This training regime is applied exclusively to Track B. 
For Track A, we train separate models on the public and hidden subsets. The rationale for this choice is that the classification tasks in Track A rely on discrete supervised labels. 
Merging the public and hidden data would result in a different training distribution and label space. 
Therefore, we maintain separate decoders to ensure that the challenge results remain a faithful reflection of the participants' development performance.

\section{Challenge Design}

The challenge utilizes a unified end-to-end generative evaluation system. Participants provide a pretrained encoder on top of which the organizers then train a lightweight projector to interface the encoder with a pretrained LLM.

\begin{figure}
    \centering
    \includegraphics[width=1.0\linewidth]{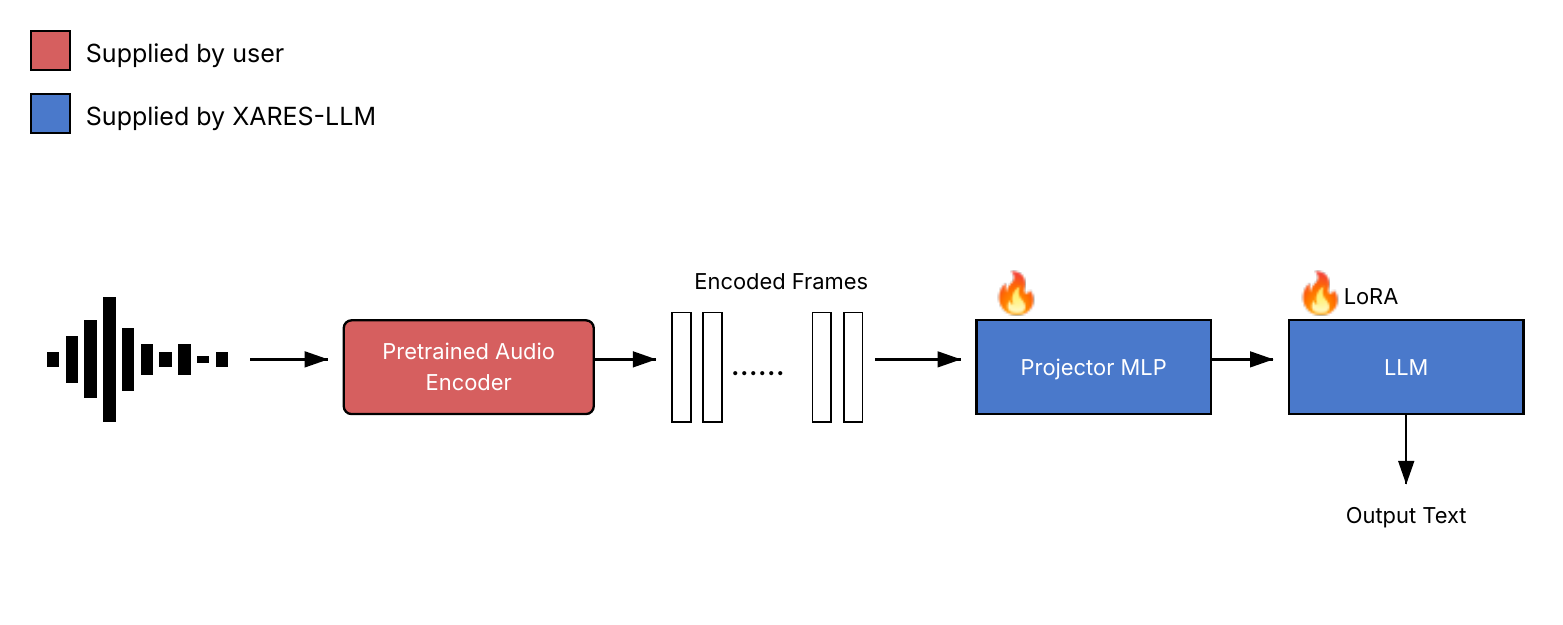}
    \caption{The XARES-LLM framework. Participants provide a frozen encoder, and the framework trains an LLM.}
    \label{fig:framework}
\end{figure}

\subsection{Evaluation System: XARES-LLM}

The organizers employ the open-source XARES-LLM framework to train a projection layer that bridges the submitted encoder with a pre-trained decoder LLM.
Specifically, we use SmolLM2-135M~\cite{allal2025smollm2} as the main decoder model, since it is small enough to allow for fast training, while ensuring that the system's performance is primarily 
\begin{table}[hb]
    \centering
    \caption{Overview of Challenge Tasks and Model Training}
    \label{tab:task_structure}
    \begin{tabular}{r|ll}
        \toprule
        {Track} & {Objective} & {LLM Decoder} \\
        \midrule
        A & \multirow{2}{*}{Classification} & \multirow{2}{*}{Dedicated}  \\
        A (Hidden) &  & \\
        \hline
        B & \multirow{2}{*}{Understanding}  & Dedicated \\
        B (Hidden) &  & Shared with Track B \\ 
        \bottomrule
    \end{tabular}
\end{table}
\begin{table}[htbp]
\centering
\caption{Track A (Classification) datasets in XARES-LLM, including mapping to Leaderboard IDs ($P_n, H_n$). Domains: \colorbox{xiaomiblue!5}{blue} (speech), \colorbox{yellow!5}{yellow} (sound), and \colorbox{xiaomired!5}{red} (music).}
\label{tab:combined_classification_datasets}
\resizebox{\linewidth}{!}{
\begin{tabular}{llllr}
\toprule
\textbf{ID} & \textbf{Dataset} & \textbf{Task Type} & \textbf{Metric} & \textbf{\#} \\
\midrule
\rowcolor{xiaomiblue!5} $P_{1}$  & ASV2015 \cite{kinnunen2018automatic} & Spoofing & Acc & 2 \\
\rowcolor{xiaomiblue!5} $P_{2}$  & CREMA-D \cite{cao2014crema} & Emotion & Acc & 5 \\
\rowcolor{xiaomiblue!5} $P_{4}$  & Fluent Spch Comm \cite{lugosch2019speech} & Intent & Acc & 248 \\
\rowcolor{xiaomiblue!5} $P_{5}$  & LibriSpeech \cite{panayotov2015librispeech} & Gender & Acc & 2 \\
\rowcolor{xiaomiblue!5} $P_{9}$  & LibriCount \cite{stoter2018libricount} & Speaker count & Acc & 11 \\
\rowcolor{xiaomiblue!5} $P_{11}$ & Speech Commands \cite{warden2018speech} & Keyword spotting & Acc & 30 \\
\rowcolor{xiaomiblue!5} $P_{13}$ & VocalSound \cite{gong_vocalsound} & Non-speech & Acc & 6 \\
\rowcolor{xiaomiblue!5} $P_{14}$ & VoxCeleb1-Bin \cite{nagrani2020voxceleb} & Speaker ID (bin) & Acc & 2 \\
\rowcolor{xiaomiblue!5} $P_{15}$ & VoxLingua33 \cite{valk2021voxlingua107} & Lang ID & Acc & 33 \\
\midrule                
\rowcolor{yellow!5} $P_{3}$  & ESC-50 \cite{piczak2015esc} & Environment & Acc & 50 \\
\rowcolor{yellow!5} $P_{6}$  & FSD50k \cite{fonseca2021fsd50k} & Sound event & mAP & 200 \\
\rowcolor{yellow!5} $P_{7}$  & FSD18-Kaggle \cite{fonseca2018general} & Sound event & mAP & 41 \\
\rowcolor{yellow!5} $P_{12}$ & UrbanSound 8k \cite{salamon2014dataset} & Urban sound & Acc & 10 \\
\midrule                
\rowcolor{xiaomired!5} $P_{5}$ & FMA Small \cite{defferrard2016fma} & Music genre & Acc & 8 \\
\rowcolor{xiaomired!5} $P_{8}$  & GTZAN Genre \cite{sturm2013gtzan} & Music genre & Acc & 10 \\
\rowcolor{xiaomired!5} $P_{10}$ & NSynth-Instr. \cite{nsynth2017} & Instruments & Acc & 11 \\
\midrule
\multicolumn{5}{c}{\textit{Hidden Evaluation Datasets}} \\
\midrule
\rowcolor{yellow!5} $H_{1}$ & FingerSnap & Fingersnap  & Acc & 2 \\
\rowcolor{yellow!5} $H_{2}$ & KeyScratching & Car Security & Acc & 2 \\
\rowcolor{yellow!5} $H_{3}$ & King-ASR-457 & Indoor Ambience & Acc & 4 \\
\rowcolor{yellow!5} $H_{4}$ & King-ASR-719 & Domestic Life & Acc & 3 \\
\rowcolor{yellow!5} $H_{5}$ & King-ASR-876 & Commercial & Acc & 3 \\
\rowcolor{yellow!5} $H_{6}$ & King-ASR-955 & Transport &  Acc & 3 \\
\bottomrule
\end{tabular}
}
\end{table}
attributed to the audio encoder’s representations rather than the LLM’s internal knowledge.

Unlike previous iterations~\cite{zhang2025xares} that used linear probing on classification tasks, XARES-LLM evaluates the encoder's performance in an end-to-end generative framework.
The XARES-LLM framework jointly optimizes two components: a lightweight projector and a LoRA-enhanced~\cite{hu2022lora} decoder LLM. 
The framework can be seen in \Cref{fig:framework}.

Ensuring a reproducible setup presents several technical challenges, primarily regarding the synchronization of random seeds. 
However, even with all seeds fixed, absolute results may vary across hardware environments due to the nondeterministic nature of graphics processing units (GPUs).
However, during development of the toolkit we experimented with three different types of available GPUs and came to the conclusion that, while absolute results vary, relative ranks remain consistent.
Consequently, all reported results are derived from our internal execution environment.
For all experiments, we utilized a default batch size of $4$, a learning rate of $10^{-4}$, and the standard number of training iterations.

\subsection{Task Categories and evaluation metrics}

The evaluation covers two primary components:
\begin{itemize}
\item Classification tasks across speech, sound and music domains with tasks such as Keyword spotting, Speaker ID, Emotion using Accuracy and mAP. For all classification tasks, we train the LLM to predict the labels directly. The datasets selected for Track A can be seen in \Cref{tab:combined_classification_datasets}.
\item Understanding tasks such as Speech Transcription (ASR) measured by inverted Word Error Rate (iWER) and Sound/Music/General Captioning measured by FENSE and DATE metrics. The datasets selected for Track B can be seen in \Cref{tab:track_b_combined}.
\end{itemize}

For all classification tasks, model performance is evaluated by directly comparing the generated output strings against the ground truth. 
To ensure a robust assessment, we apply a text normalization scheme that removes hyphens and redundant whitespace. 
For multi-label tasks (FSD50k, FSD18-Kaggle) requiring simultaneous predictions, a custom separator (;) is employed to delimit the labels.
\begin{table}[htbp]
\centering
\caption{Track B (Understanding) datasets in XARES-LLM, including hidden evaluation tasks. Public versions are hosted on Huggingface. Domains: \colorbox{xiaomiblue!5}{blue} (speech), \colorbox{yellow!5}{yellow} (sound), and \colorbox{xiaomired!5}{red} (music).}
\label{tab:track_b_combined}
\resizebox{\linewidth}{!}{
\begin{tabular}{llll}

\toprule
{ID} & {Dataset} & {Task Type} & {Metric} \\
\midrule
\rowcolor{xiaomiblue!5} $U_1$ & AISHELL-1-100h \cite{aishell_2017} & ASR & iCER \\
\rowcolor{yellow!5} $U_2$ & Clotho \cite{drossos2020clotho} & Sound Caption & FENSE \\
\rowcolor{xiaomiblue!5} $U_3$ & LibriSpeech-100h \cite{panayotov2015librispeech} & ASR & iWER \\
\rowcolor{white} $U_4$ & MECAT~\cite{mecat2025} & General Caption & DATE \\
\rowcolor{xiaomired!5} $U_5$ & The Song Describer Dataset \cite{manco2023thesong} & Music Caption & FENSE \\

\midrule
\multicolumn{4}{c}{\textit{Hidden Evaluation Datasets}} \\
\midrule
\rowcolor{xiaomiblue!5} $E_1$ & AISHELL-6 \cite{aishell6} & ASR & iCER \\
\rowcolor{xiaomiblue!5} $E_2$ & LibriHeavy \cite{libriheavy} & ASR & iWER \\
\rowcolor{xiaomired!5} $E_3$ & MusicCaps \cite{manco2023thesong} & Music Caption & FENSE \\
\rowcolor{yellow!5} $E_4$ & TACOS \cite{tacos} & Sound Caption & FENSE \\
\bottomrule
\end{tabular}
}
\end{table}

\subsection{Scoring}

Following the task structure outlined in \Cref{tab:task_structure}, we conducted three distinct experiments, resulting in three specialized LLMs. 
The results for Track B (Hidden) were derived directly from the model trained on Track B data, serving as a benchmark for out-of-domain evaluation.

The final scores are computed as:
$$\text{Score}_i = \text{Avg}(\text{Track}_i) + \text{Avg}(\text{Track}_{i,\text{Hidden}}), \quad i \in \{A, B\}.$$

To ensure the reproducibility and robustness of our benchmarks, all tests were executed across multiple hardware configurations ($m \in M$, representing distinct machines/GPUs). 
$$\text{Score}_i = \text{Avg}(\max_{m \in M} \text{Track}_{i,m}) + \text{Avg}(\max_{m \in M} \text{Track}_{i,\text{Hidden},m}).$$
We observed a high degree of consistency across different machines, with a relative score discrepancy of less than 1\% for the majority of models, confirming that the results are stable and comparable across environments.

\begin{table*}[htbp]
\centering
\caption{Track A leader leaderboard results. Public ($P$) and Hidden ($H$) metrics  are provided.}
\label{tab:comprehensive_leaderboard_a}
\footnotesize 
\setlength{\tabcolsep}{3pt} 
\resizebox{\textwidth}{!}{%
\begin{tabular}{@{}l c ccccccccccccccc cccccc@{}}
\toprule
\multirow{2}{*}{\textbf{Team}} & \multirow{2}{*}{\textbf{Avg.}} & \multicolumn{15}{c}{\textbf{Public Test Results ($P_n$)}} & \multicolumn{6}{c}{\textbf{Hidden Test Results ($H_n$)}} \\
\cmidrule(lr){3-17} \cmidrule(lr){18-23}
& & $P_1$ & $P_2$ & $P_3$ & $P_4$ & $P_5$ & $P_6$ & $P_7$ & $P_8$ & $P_9$ & $P_{10}$ & $P_{11}$ & $P_{12}$ & $P_{13}$ & $P_{14}$ & $P_{15}$ & $H_1$ & $H_2$ & $H_3$ & $H_4$ & $H_5$ & $H_6$ \\
\midrule
THUVoice & \textbf{91.2} & \textbf{99.0} & \textbf{87.9} & 91.3 & 99.4 & \textbf{95.9} & 32.1 & 88.4 & 90.9 & 47.9 & \textbf{78.8} & \textbf{93.5} & 88.9 & 93.9 & 97.2 & 93.9 & 87.5 & \textbf{99.9} & 98.5 & 98.1 & \textbf{100.0} & \textbf{99.0} \\
THU-HSCI-2 & 90.8 & 98.6 & 86.0 & 92.3 & 99.4 & 89.1 & \textbf{33.9} & \textbf{89.1} & \textbf{93.9} & 50.6 & 78.7 & 82.6 & \textbf{89.6} & \textbf{94.5} & 97.7 & 87.3 & 87.5 & 99.7 & {99.5} & \textbf{98.8} & \textbf{100.0} & 83.0 \\
trans-encoder & 90.0 & 98.3 & 87.3 & 92.3 & \textbf{99.6} & 73.1 & 32.4 & 87.6 & 92.9 & \textbf{50.9} & 75.6 & 84.6 & 88.1 & 93.9 & \textbf{98.5} & 94.1 & 87.5 & 95.7 & {99.5} & 98.4 & \textbf{100.0} & 98.7 \\
IASP Lab & 89.4 & 99.0 & 85.2 & 91.5 & 99.2 & 71.6 & 30.9 & 85.9 & 89.9 & 49.3 & 75.3 & 81.0 & 84.8 & 94.0 & 98.3 & \textbf{97.5} & 86.4 & 99.4 & {99.5} & 97.4 & \textbf{100.0} & 96.7 \\
MIT SLS & 88.4 & 96.7 & 69.0 & 91.3 & 99.5 & 64.4 & 27.5 & 84.2 & 90.9 & 43.6 & 72.8 & 79.1 & 88.1 & 93.7 & 97.1 & 95.5 & \textbf{87.5} & 99.8 & 99.0 & 98.0 & \textbf{100.0} & 98.7 \\
EncodexPOL & 88.1 & 98.4 & 83.1 & \textbf{93.5} & 99.5 & 87.0 & 30.8 & 85.4 & 84.9 & 45.6 & 75.4 & 77.4 & 88.3 & 94.2 & 97.4 & 89.8 & 87.3 & 94.5 & 98.5 & 96.6 & \textbf{100.0} & 88.0 \\
Fusion SUMMON & 86.5 & 98.8 & 70.4 & 82.0 & 98.6 & 62.6 & 17.4 & 71.9 & 82.8 & 48.3 & 71.6 & 79.1 & 86.6 & 92.3 & 97.5 & 96.7 & 86.7 & 96.9 & {99.5} & 97.7 & \textbf{100.0} & 94.3 \\
LvDouSha & 84.2 & 98.3 & 68.0 & 72.0 & 93.9 & 59.0 & 12.3 & 66.7 & 83.8 & 45.9 & 68.0 & 74.7 & 80.1 & 91.2 & 96.4 & 97.1 & 86.5 & 99.5 & 98.0 & 94.5 & \textbf{100.0} & 88.7 \\
Dark Reunion & 83.1 & 98.7 & 67.8 & 86.5 & 99.3 & 61.6 & 16.0 & 61.5 & 89.9 & 54.3 & 74.4 & 78.1 & 85.8 & 93.7 & 98.0 & 94.5 & 84.7 & 94.5 & 89.5 & 96.0 & 93.7 & 75.3 \\
NTUST\_NLPLab & 82.3 & 95.9 & 67.9 & 80.0 & 98.4 & 65.0 & 18.0 & 72.4 & 85.9 & 48.9 & 70.0 & 69.6 & 84.8 & 93.2 & 91.9 & 91.1 & 86.4 & 94.8 & 92.0 & 93.2 & 99.7 & 68.3 \\
CORE & 79.8 & 98.2 & 56.5 & 61.0 & 98.7 & 57.8 & 8.5 & 45.7 & 73.7 & 47.1 & 66.8 & 70.6 & 78.1 & 90.7 & 89.9 & 69.6 & 86.4 & 85.1 & 96.5 & 93.9 & \textbf{100.0} & 90.3 \\
DKU-WHU & 78.9 & 97.4 & 65.0 & 71.3 & 93.5 & 60.6 & 10.5 & 53.0 & 72.7 & 44.8 & 69.9 & 80.4 & 82.0 & 89.8 & 85.7 & 86.3 & 85.7 & 94.6 & 86.0 & 92.9 & 94.0 & 68.3 \\
WaWu & 78.9 & 96.2 & 51.1 & 69.8 & 99.0 & 63.3 & 13.1 & 49.7 & 71.7 & 48.8 & 57.6 & 72.3 & 76.8 & 91.5 & 83.6 & 82.9 & 85.9 & 98.6 & 95.0 & 92.5 & 97.0 & 66.7 \\
audio eevee & 78.7 & 98.5 & 70.7 & 81.8 & 99.4 & 61.9 & 17.8 & 68.3 & 87.9 & 55.3 & 74.8 & 78.2 & 86.6 & 93.6 & 97.7 & 93.2 & 82.2 & 91.4 & 73.5 & 87.2 & 74.7 & 69.7 \\
NTU\_BIO & 75.9 & 92.8 & 46.6 & 60.8 & 67.9 & 60.9 & 9.1 & 55.6 & 78.8 & 34.4 & 65.9 & 64.7 & 71.7 & 88.1 & 69.6 & 79.9 & 85.3 & 87.8 & 98.0 & 85.9 & 99.7 & 75.0 \\
SAMoVA & 73.5 & 96.5 & 43.0 & 64.8 & 19.9 & 61.0 & 7.9 & 45.4 & 71.7 & 25.2 & 72.5 & 12.3 & 76.1 & 84.0 & 79.7 & 5.3 & 86.6 & 96.4 & \textbf{100.0} & 97.5 & \textbf{100.0} & 95.0 \\
Echo bridge & 72.8 & 91.1 & 52.8 & 87.5 & 77.4 & 63.8 & 20.4 & 69.5 & 84.9 & 35.6 & 69.8 & 61.3 & 82.6 & 91.0 & 89.6 & 77.5 & 83.7 & 92.9 & 62.5 & 91.1 & 63.3 & 58.7 \\
XJTLU & 72.8 & 95.1 & 28.4 & 75.0 & 98.4 & 53.3 & 10.9 & 61.6 & 80.8 & 22.4 & 51.8 & 61.0 & 86.1 & 90.5 & 96.9 & 12.1 & 82.7 & 86.5 & 77.5 & 87.4 & 93.0 & 77.0 \\
JustForFun & 69.3 & 97.8 & 39.2 & 39.3 & 98.9 & 51.8 & 3.9 & 24.5 & 66.7 & 35.3 & 54.5 & 70.5 & 60.5 & 88.1 & 39.6 & 50.5 & 77.0 & 95.3 & 88.5 & 81.9 & 92.0 & 68.0 \\
WindSpeak & 67.4 & 96.8 & 74.2 & 76.5 & 99.2 & 53.9 & 14.1 & 67.4 & 77.8 & 50.4 & 75.4 & 78.9 & 80.3 & 93.4 & 98.4 & 89.3 & 85.6 & 92.8 & 28.0 & 69.4 & 41.3 & 41.3 \\
RBG-AI & 39.5 & 95.1 & 33.4 & 14.3 & 11.0 & 40.5 & 2.4 & 5.7 & 47.5 & 28.0 & 29.4 & 9.3 & 30.0 & 50.3 & 47.1 & 6.4 & 60.0 & 72.2 & 51.0 & 38.3 & 39.0 & 33.3 \\
Pinch & 38.6 & 95.1 & 35.4 & 6.3 & 8.4 & 19.6 & 2.0 & 4.5 & 18.2 & 16.7 & 26.0 & 4.6 & 18.4 & 27.8 & 26.0 & 3.7 & 60.0 & 72.2 & 56.5 & 38.5 & 67.7 & 43.0 \\
AICIS & 28.6 & 93.0 & 25.4 & 2.0 & 10.0 & 35.3 & 2.5 & 7.7 & 43.4 & 8.2 & 11.6 & 3.8 & 11.8 & 31.8 & 38.9 & 6.0 & 60.0 & 72.2 & 28.0 & 7.9 & 27.0 & 16.0 \\
\bottomrule
\end{tabular}%
}

\vspace{2em}
\centering
\caption{Leaderboard results for Track B of the challenge. Teams are ranked by their final average score across public and hidden test sets. Tasks denoted with $^*$ are hidden.}
\label{tab:leaderboard_track_b}
\resizebox{0.60\linewidth}{!}{%
\begin{tabular}{@{}l c c c c c c c c c c@{}}
\toprule
\multirow{2}{*}{\textbf{Team}} & \multirow{2}{*}{\textbf{Avg.}} & \multicolumn{5}{c}{\textbf{Public Test Results}} & \multicolumn{4}{c}{\textbf{Hidden Test Results}} \\
\cmidrule(lr){3-7} \cmidrule(lr){8-11}
& & $U_1$ & $U_2$ & $U_3$ & $U_4$ & $U_5$ & $E_1$* & $E_2$* & $E_3$* & $E_4$* \\
\midrule
trans-encoder & \textbf{65.9} & \textbf{70.2} & 61.6 & 90.0 & 49.0 & \textbf{93.4} & 67.5 & \textbf{51.0} & 71.7 & 82.5 \\
WaWu          & 65.2 & 68.6 & \textbf{61.9} & 89.4 & 42.6 & 92.6 & 68.1 & 50.1 & 78.9 & 89.1 \\
THU-HSCI-2    & 65.0 & 69.6 & 60.3 & 89.8 & 49.1 & 92.4 & 68.1 & 48.7 & 67.2 & 79.4 \\
THUVoice      & 63.0 & 70.1 & 55.8 & \textbf{90.9} & 47.9 & 92.5 & \textbf{69.7} & 49.7 & 58.3 & 73.7 \\
EncodexPOL    & 62.3 & 66.3 & 58.3 & 86.6 & 47.5 & 83.3 & 67.2 & 46.9 & 68.2 & 75.3 \\
NTUST\_NLPLab & 62.0 & 66.1 & 57.9 & 86.0 & 42.0 & 87.5 & 64.5 & 50.6 & 68.8 & 81.9 \\
MIT SLS       & 61.5 & 66.0 & 57.0 & 75.8 & 47.0 & 92.2 & 66.8 & 48.2 & 54.6 & 87.4 \\
audio eevee   & 48.6 & 54.8 & 42.3 & 44.5 & 43.2 & 76.4 & 65.3 & 44.7 & 21.8 & 68.6 \\
Dark Reunion  & 47.1 & 52.3 & 41.9 & 32.4 & 42.8 & 74.0 & 64.8 & 47.7 & 18.0 & 66.3 \\
IASP Lab      & 46.8 & 51.9 & 41.8 & 47.8 & \textbf{49.4} & 44.8 & 66.4 & 51.0 & 35.0 & 41.9 \\
Fusion SUMMON  & 45.1 & 49.9 & 40.3 & 46.5 & 43.1 & 44.5 & 66.3 & 49.2 & 36.3 & 43.2 \\
LvDouSha      & 44.5 & 49.2 & 39.8 & 47.2 & 39.4 & 45.6 & 65.3 & 48.6 & 36.5 & 44.5 \\
XJTLU         & 43.0 & 24.2 & 61.8 & 0.0  & 33.6 & 0.0  & 42.0 & 45.7 & \textbf{79.4} & \textbf{89.5} \\
NTU\_BIO      & 42.0 & 47.1 & 36.9 & 38.4 & 35.8 & 52.9 & 61.8 & 46.4 & 26.7 & 44.4 \\
Echo bridge   & 40.2 & 45.2 & 35.3 & 31.7 & 44.7 & 36.7 & 64.1 & 48.9 & 21.1 & 35.5 \\
WindSpeak     & 40.0 & 45.6 & 34.4 & 24.7 & 36.0 & 59.1 & 62.1 & 46.1 & 14.6 & 50.5 \\
DKU-WHU       & 39.5 & 44.6 & 34.5 & 35.8 & 34.4 & 39.5 & 62.7 & 50.6 & 25.2 & 37.8 \\
JustForFun    & 33.0 & 29.7 & 36.3 & 1.2  & 24.5 & 32.8 & 53.7 & 36.5 & 4.1  & 72.9 \\
CORE          & 30.3 & 34.3 & 26.3 & 0.0  & 35.5 & 25.7 & 60.3 & 50.1 & 0.0  & 31.3 \\
SAMoVA        & 22.8 & 27.9 & 17.7 & 0.0  & 37.6 & 0.0  & 55.2 & 46.8 & 2.9  & 0.0  \\
AICIS        & 17.6 & 19.6 & 15.6 & 0.0  & 12.9 & 0.0  & 45.6 & 39.6 & 4.1  & 0.0  \\
Pinch         & 16.3 & 19.1 & 13.5 & 0.0  & 11.4 & 0.0  & 37.7 & 46.5 & 4.1  & 0.0  \\
RBG-AI        & 15.4 & 16.8 & 13.9 & 0.0  & 8.8  & 0.0  & 32.6 & 42.8 & 2.4  & 0.4  \\
\bottomrule
\end{tabular}%
}
\end{table*}
\begin{table*}[t]
\centering
\caption{Challenge submissions, their approach and (if applicable) the utilized pre-trained encoders. Methods utilizing audio encoders trained with a LALM are highlighted in \colorbox{xiaomimedgray!10}{grey}.}
\label{tab:team_comparison}
\resizebox{0.95\linewidth}{!}{
\begin{tabular}{l|l|l|c|c}
\toprule
{Team} & {Strategy} & {Pre-trained Encoders} & {Track A Rank} & {Track B Rank} \\
\midrule
\rowcolor[gray]{0.9} THU-Voice & Mixture-of-Experts (MoE) & Audio-Flamingo 3, Qwen2-Audio & 1 & 4 \\ 
\rowcolor[gray]{0.9} THU-HSCI-2 & Dual-Model Fusion & Qwen2-Audio, Audio-Flamingo 3 & 2 & 3 \\ 
\rowcolor[gray]{0.9} trans-encoder & Hierarchical Extraction & StepAudio 2 & 3 & 1 \\ 
\rowcolor[gray]{0.9} IASP Lab & Dual-Branch Time Fusion & Qwen2-Audio, Whisper & 4 & 10 \\ 
\rowcolor[gray]{0.9} EncodexPOL & Audio-text alignment & Audio-Flamingo 3, Whisper & 6 & 5 \\ 
\rowcolor[gray]{0.9} NTU\_BIO & Gated Feature Fusion & Qwen3-ASR & 15 & 14 \\ 
MIT SLS & Domain Teacher Distillation & USAD & 5 & 7 \\ 
WaWu & Dual-Branch Gated Fusion & Whisper, WavLM-Base-Plus & 13 & 2 \\ 
LvDouSha & Prefix fusion & Whisper & 8 & 12 \\
Fusion SUMMON & Tri audio encoder & Dasheng, Whisper, mHubert & 7 & 11 \\ 
DKU-WHU & Feature Fusion & Dasheng, Whisper & 12 & 17 \\ 
Echo bridge & Feature Fusion & Whisper, CED-mini & 17 & 15 \\ 
XJTLU & MoE Alignment & Whisper, BEATs & 18 & 13 \\ 
JustForFun & Supervised Classification & WavLM-base-plus & 19 & 18 \\ 
RBG-AI & Gated Feature Fusion & ECAPA-TDNN, Wav2Vec2-BERT & 21 & 23 \\ 
CORE & Spectro-Temporal Predictive Modeling & - & 11 & 19 \\ 
SAMoVA & ViT based BEST-RQ & - & 16 & 20 \\ 
Pinch & JEPA Audio Encoder & - & 22 & 22 \\ 
AICIS & Convolutional-Transformer hybrid & - & 23 & 21 \\
\bottomrule
\end{tabular}
}
\end{table*}

\section{Results}

The performance for Tracks A and B is detailed in \Cref{tab:comprehensive_leaderboard_a} and \Cref{tab:leaderboard_track_b}, respectively. 

In Track A, THUVoice achieved the top position with an average score of 91.2\%, narrowly surpassing THU-HSCI-2 (90.8\%) and trans-encoder (90.0\%). 
THUVoice secured the overall lead, driven by its superior robustness on the hidden test sets.
It achieved 99.0\% on ASV2015 and 97.2\% on VoxCeleb1-Bin, suggesting a superior capability in extracting speaker-specific embeddings compared to other approaches.
THU-HSCI-2 followed closely in second place with an average of 90.8\%, distinguishing itself as the premier model for environmental and musical audio analysis. 
It recorded the highest scores across several public benchmarks, including FSD50k at 33.9\%, FSD18-Kaggle at 89.1\%, and GTZAN Genre at 93.9\%. 
These results indicate that THU-HSCI-2 possesses a superior ability to extract features from acoustic scenes.
Lastly, trans-encoder achieves the best performance on Fluent Speech commands, LibriCount and VoxCeleb1-Bin, signaling its strong performance for intent classification, speaker counting and speaker verification.

In Track B, trans-encoder secures the highest average score of 65.9\%, followed by WaWu and THU-HSCI-2.
WaWu's second place performance with a score of 65.2\%, shows competitive performance on the Clotho dataset, while THU-HSCI-2 maintained a top-three standing with an average of 65.0\%.

\begin{figure}
    \centering
    \includegraphics[width=1.05\linewidth]{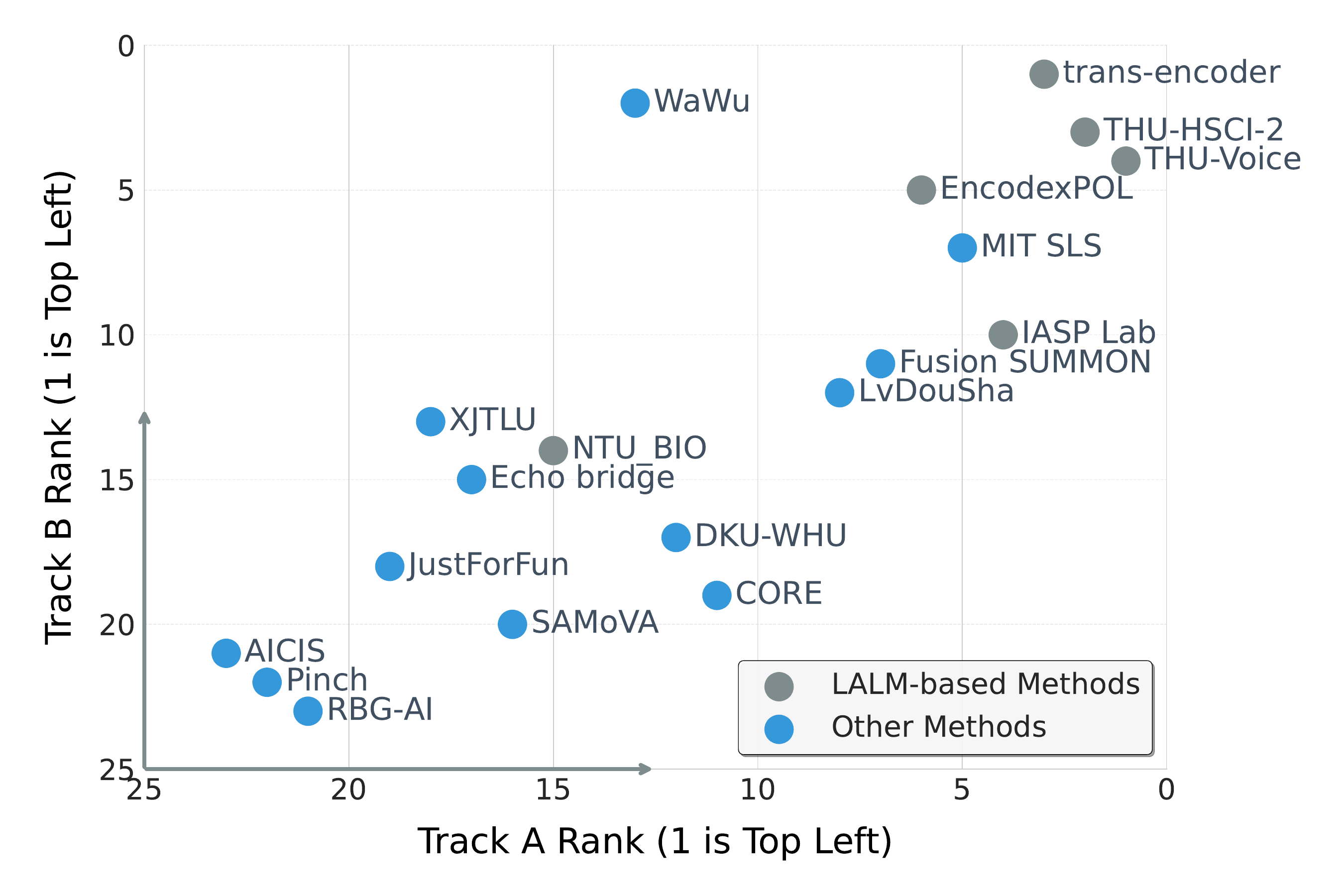}
    \caption{Comparison of team rankings across Track A and Track B.}
    \label{fig:rank_comparison}
\end{figure}

A short description of the utilized approaches is given in \Cref{tab:team_comparison} and visualized in \Cref{fig:rank_comparison}.
A trend among the top-tier submissions was the use of proprietary audio encoders, such as Qwen2-Audio~\cite{chu2024qwen2} and Step-Audio 2~\cite{wu2025step} and Audio-Flamingo 3~\cite{audioflamingo3}.
These models provide a significant performance boost due to their extensive pre-training on diverse audio-text datasets using an LLM decoder.
While experiments involved a wide variety of different pretrained models, the results of the challenge indicate a specific pattern.
First, utilizing a pretrained audio encoder is fundamental and greatly improves performance compared to training from scratch.
Second, feature fusion strategies generally boost performance across the majority of participants.
Most significantly, utilizing an audio encoder pre-trained within an LALM framework appears to be  the most effective approach for creating general, multi-domain audio encoders.

\section{Discussion and Conclusion}

This paper introduced XARES-LLM, a streamlined framework designed for the training and evaluation of Large Audio Language Models in the context of general audio understanding. 
By leveraging a frozen pretrained encoder, a trainable projector, and parameter-efficient fine-tuning via LoRA, XARES-LLM provides a robust architecture for bridging the gap between raw audio signals and textual reasoning.

The Interspeech 2026 Audio Encoder Capability Challenge successfully utilized this framework to benchmark submissions from 22 teams worldwide. 
The results of this challenge demonstrate a measurable performance advantage for audio encoders that utilize Large Audio-Language Model (LALM) alignment. 
In Track A, the top three teams (THUVoice, THU-HSCI-2, and trans-encoder) all employed encoders pre-trained on diverse audio-text datasets via LLM decoders. 
This suggests that audio-text alignment with LALMs results in a more robust feature space than that produced by standard self-supervised learning (SSL) or supervised classification on single-domain datasets.
The competition results demonstrate that while current state-of-the-art encoders achieve high accuracy on established public datasets, challenges remain in maintaining performance across out-of-domain and hidden test sets. 
Ultimately, these findings confirm that the integration of a pre-trained LLM during the audio encoder’s training phase is the most effective current methodology for developing general-purpose audio representations. 
Alignment with an LLM appears to provide the necessary robustness to handle varied tasks within a single, unified framework.



\newpage
\bibliographystyle{IEEEtran}
\bibliography{mybib}

\end{document}